\documentclass[conference]{IEEEtran}
\IEEEoverridecommandlockouts
\usepackage{cite}
\usepackage{amsmath,amssymb,amsfonts}
\usepackage{algorithmic}
\usepackage{graphicx}
\usepackage{textcomp}
\usepackage{xcolor}
\usepackage{subcaption}
\def\BibTeX{{\rm B\kern-.05em{\sc i\kern-.025em b}\kern-.08em
    T\kern-.1667em\lower.7ex\hbox{E}\kern-.125emX}}
\begin{document}

\title{CryptoGuard: An AI-Based Cryptojacking Detection Dashboard Prototype\\}

\author{\IEEEauthorblockN{Amitabh Chakravorty}
\IEEEauthorblockA{\textit{School of Information Technology} \\
\textit{University of Cincinnati}\\
Cincinnati, Ohio, USA \\
chakraa4@mail.uc.edu}
\and
\IEEEauthorblockN{Jess Kropczynski}
\IEEEauthorblockA{\textit{School of Information Technology} \\
\textit{University of Cincinnati}\\
Cincinnati, Ohio, USA \\
kropczjn@ucmail.uc.edu}
\and
\IEEEauthorblockN{Nelly Elsayed}
\IEEEauthorblockA{\textit{School of Information Technology} \\
\textit{University of Cincinnati}\\
Cincinnati, Ohio, USA \\
elsayeny@ucmail.uc.edu}
}

\maketitle

\begin{abstract}
With the widespread adoption of cryptocurrencies, cryptojacking has become a significant security threat to crypto wallet users. This paper presents a front-end prototype of an AI-powered security dashboard, namely, CryptoGuard. Developed through a user-centered design process, the prototype was constructed as a high-fidelity, click-through model from Figma mockups to simulate key user interactions. It is designed to assist users in monitoring their login and transaction activity, identifying any suspicious behavior, and enabling them to take action directly within the wallet interface. The dashboard is designed for a general audience, prioritizing an intuitive user experience for non-technical individuals. Although its AI functionality is conceptual, the prototype demonstrates features like visual alerts and reporting. This work is positioned explicitly as a design concept, bridging cryptojacking detection research with human-centered interface design. This paper also demonstrates how usability heuristics can directly inform a tool's ability to support rapid and confident decision-making under real-world threats. This paper argues that practical security tools require not only robust backend functionality but also a user-centric design that communicates risk and empowers users to take meaningful action.
\end{abstract}


\section{Introduction}

Consider a scenario in which an individual notices that their mobile device is generating abnormal heat despite the absence of any visible or background activities. Several users of cryptocurrency and digital wallet applications have reported such incidents. An apparent explanation for such cases of cryptojacking, a rapidly emerging type of cyberattack, is that malicious actors exploit a victim’s computing resources without authorization \cite{nahmias2019}. 

In this case, the victim can be an individual or an entire organization, and the computing resources may include mobile devices, tablets, laptops, desktops, cloud or on-premises servers, virtual machines or containers, IoT devices, and other internet-enabled devices. This usually happens when malicious scripts are injected into web pages, mobile applications, or systems. Once the device is turned on, the script quietly utilizes some central processing unit (CPU) resources and gradually slows the device down without obvious signs for a long time \cite{nahmias2019}. Unlike ransomware or phishing attacks, which create immediate chaos, cryptojacking usually runs quietly in the background, making it more difficult to detect and deal with \cite{nahmias2019}. As mentioned by Moreno-Sancho et al. \cite{moreno2024}, cryptojacking attacks target a variety of systems, like network infrastructures and digital wallets. Wang et al. \cite{wang2024} also note that attackers are employing advanced deep learning techniques to evade detection. According to recent academic studies, it is evident that the rise of mobile and cloud-based applications for managing cryptocurrencies has made these platforms more susceptible to risks. Thus, traditional cybersecurity tools tend to be more reactive, which makes it challenging for them to detect the more subtle signs of attacks promptly \cite{danesh2024}.

To address these emerging threats, this paper introduces CryptoGuard, a conceptual, high-fidelity front-end prototype of an AI-powered security dashboard designed for integration within a cryptocurrency wallet application. Focusing on login and transaction activities, the dashboard would offer real-time monitoring of account activity. It utilizes a simulated Support Vector Machine (SVM) model, which identifies abnormal activity trends that may indicate someone is attempting to use others' computing resources for illicit cryptocurrency mining. If a risk is detected, the AI detection system will send a warning and provide the user with easy steps to take. Among these steps are notifying the wallet provider of the event and selecting actions such as applying an account freeze or requesting transaction information \cite{rani2024}.


The proposed CryptoGuard is distinguished by its focus on transparency and user responsiveness. The system's dashboard is designed to provide intuitive insights, enabling even non-technical users to identify potential anomalies readily.
Summaries of login and transaction activity, paired with attention-grabbing alerts and built-in reporting links, are all intended to help users act swiftly and confidently. As Yu et al.~\cite{yu2024} noted, users often balance convenience with safety, and visual transparency is key to building that trust.


The primary contribution of this paper is to provide a detailed exploration of the user-centered design process behind the CryptoGuard prototype. This work documents the design rationale, information architecture, and key UI/UX decisions made to meet the security needs of non-technical users in practical scenarios. By articulating the thought process behind the prototype's development, this paper provides a foundational framework for future research, including subsequent iterations involving usability testing and the development of a functional AI detection model. 

This paper is deliberately positioned as an early-stage design concept.
The choice to emphasize design over implementation stems from a gap in current crypto wallet security solutions: most focus heavily on detection mechanics while overlooking usability for non-technical users. Our contribution centers on showing how a clear, accessible, and trust-building interface can be designed for a complex security function, even before the AI backend is implemented.

\section{Literature Review}

A unique characteristic of cryptojacking is its high degree of adaptability for different systems. Danesh, Karimi, and Arasteh \cite{danesh2024} noted that cybercriminals continually refine their techniques to exploit network data while evading conventional security alerts. Their study indicated that NetFlow-based machine learning models are capable of identifying abnormal patterns; however, it is essential to integrate them thoughtfully into existing platforms. Al-Fawareh et al. \cite{alfawareh2024} discussed how botnet-based cryptojacking can be executed across IoT networks due to its adaptability.

Even with the current progress in cryptojacking detection, most available tools are still reactive. Many systems only react to cryptojacking incidents after they have already been affected. This reactive approach often fails to keep pace with the speed required for real-time user environments, such as mobile crypto wallets. The study by Sanda, Pavlidis, and Polatidis \cite{sanda2023} demonstrates that stopping host-based cryptojacking is challenging, especially when there are no tools that users can easily identify or utilize to detect it. Similarly, Cui and Liu \cite{cui2024} discuss methods of in-browser cryptojacking and emphasize the importance of having lightweight and user-friendly detection systems that do not depend too much on backend monitoring.


AI models provide a more proactive approach to identifying cryptojacking. Models such as support vector machines and optimal neural networks are somewhat successful in detecting anomalies in transaction data, login history, and CPU usage \cite{rani2024}. These technologies enable systems to detect risks early and reduce false positives. Putting these models into practice in real-world situations, nonetheless, calls for both precise detection and interfaces that enable users to grasp what is occurring and how to respond.

Usability is a powerful aspect in cybersecurity designs. Most crypto or digital wallet users are not cybersecurity professionals and may not grasp technical notifications. The research paper by Yu et al. \cite{yu2024} reveals that users typically choose wallets based on a combination of perceived security and simplicity. They prefer technologies that allow them to understand warnings without needing to know how the detection works. According to Houy, Schmid, and Bartel \cite{houy2023}, usability issues may prevent users from fully benefiting from the security features of a digital wallet.

CryptoGuard was designed to sit at this intersection, combining AI-powered anomaly detection with a straightforward, visually driven interface. The design focuses on providing just enough information at the right moment, using color-coded alerts, summaries, and simple response options to help users make sense of what is going on. Our goal was not just to build a secure system but one that feels accessible and reassuring to the everyday user. According to Nielsen \cite{nielsen1993}, even a system that works well can still fail if the user interface does not allow for intuitive interaction.

Ultimately, this literature review suggests that cryptojacking is not merely a technical issue, but also a design problem. To build a really effective defense, the solution needs to be fast, innovative, and user-friendly. These ideas played a central role in how we approached the design of CryptoGuard, not just in terms of what it does, but in how it communicates, guides, and supports the person using it. 

Upon reviewing current wallet security tools, such as MetaMask, Trust Wallet, and Coinbase Wallet, we observed a common trend: strong backend detection mechanisms but minimal effort in making alerts immediately understandable to everyday users. Alerts often appear as short technical warnings without contextual guidance. In contrast, CryptoGuard’s concept pairs detection outputs with step-by-step actionable paths. This approach is inspired by prior findings that human factors can significantly impact the adoption of security measures. This contrast highlights the value of our contribution: bringing a human-centered lens to a space that technical considerations have mainly shaped.

\section{Methods}
This study employed a user-centered design approach to develop a conceptual prototype of CryptoGuard, an AI-powered dashboard for detecting cryptojacking. The primary objective of this work was to explore the user interface (UI) and user experience (UX) design for a security tool targeting non-technical cryptocurrency wallet users. The research was focused on creating a high-fidelity interactive prototype for future usability testing. 

\subsection{Prototype Overview}
CryptoGuard is a proposed AI-powered security dashboard designed to be integrated into any digital or crypto wallet application. Rather than being tailored to one specific platform, it was developed as a flexible and standalone module that could be embedded within existing wallet interfaces. By providing users with real-time access to their account activities, CryptoGuard would enable them to identify and respond to cryptojacking attempts. Although the AI detection capability is not yet functional, the interface visually simulates how alerts would be presented based on suspicious activity patterns.

\subsection{Prototype Development and Tools}
The design process began with the creation of high-fidelity mockups in Figma, a collaborative design tool for interface development. This stage focused on developing key screens of the CryptoGuard dashboard, including views for alert notifications, system status, and the user workflow for reporting suspicious activity. To simulate interactivity, these static mockups were then exported and assembled into a click-through prototype using Microsoft PowerPoint. This method enabled the creation of navigational pathways and simulated user interactions with dashboard features, such as acknowledging an alert or initiating a report.

\subsection{Design Principles and Justification}
The prototype's design was informed by established principles of user-centered design and usability heuristics, prioritizing clarity and ease of use for a non-expert audience \cite{nielsen1993}. Key considerations included a logical information architecture, intuitive labeling, and the visual design of alert systems to convey risk without causing undue alarm. To represent the functionality of the conceptual AI engine, the prototype includes simulated yet plausible threat scenarios, such as flagging a suspicious login or highlighting an unusual transaction. The design of the reporting workflow was specifically structured to be a low-friction process, guiding the user toward meaningful action. 

While functional integration remains a future step, this design stage deliberately focused on the "last mile" of security, the point where technical detection meets human decision-making. In doing so, the concept bridges the gap between cryptojacking detection research and the actual needs of non-expert users.

The conceptual AI engine was represented using a Support Vector Machine (SVM), chosen during the design phase for its strong performance in anomaly detection and outlier analysis \cite{rani2024}. Although the AI model has not been implemented yet, the prototype visually simulates the type of alerts such a model might generate in a live system.

\section{Results}
The resulting artifact of this study is a high-fidelity conceptual prototype of CryptoGuard, developed through a user-centered design methodology. This prototype represents the outcome of applying usability heuristics and iterative design strategies to a cybersecurity use case, specifically the mitigation of cryptojacking.

The prototype contains four main screens, each representing a core feature of the CryptoGuard dashboard. As shown in Fig. 1(a), the main dashboard presents a summary of recent logins and transactions. This screen displays the alert notifications generated by the conceptual AI system, along with the summaries. These alerts highlight suspicious activity, such as access from an unfamiliar device or a large and unexpected transfer.

\begin{figure*}[htbp]
\centering

\begin{minipage}[t]{0.48\textwidth}
  \centering
  \includegraphics[width=2.80\linewidth,height=8.8cm,keepaspectratio]{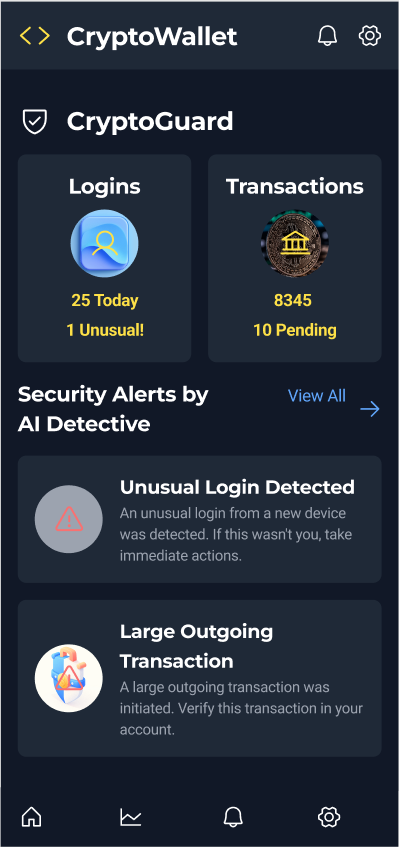}
  \caption*{(a) Home Screen}
  \vspace{0.75cm}
  \includegraphics[width=2.80\linewidth,height=8.8cm,keepaspectratio]{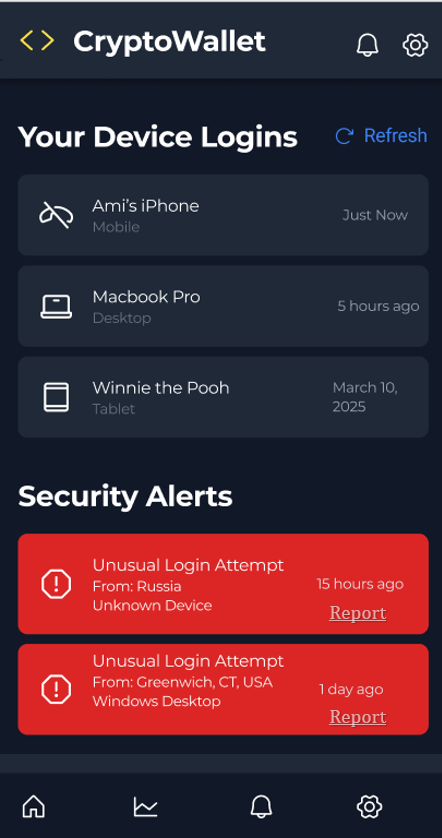}
  \caption*{(b) Login Screen}
\end{minipage}
\hspace{0.02\textwidth}
\begin{minipage}[t]{0.48\textwidth}
  \centering
  \includegraphics[width=2.80\linewidth,height=8.8cm,keepaspectratio]{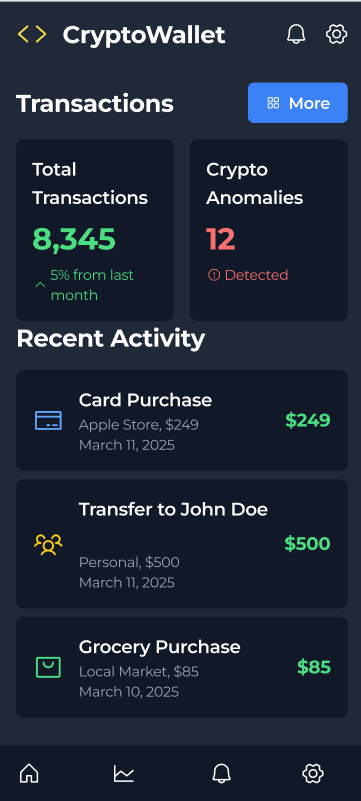}
  \caption*{(c) Transactions Screen}
  \vspace{0.75cm}
  \includegraphics[width=2.80\linewidth,height=8.8cm,keepaspectratio]{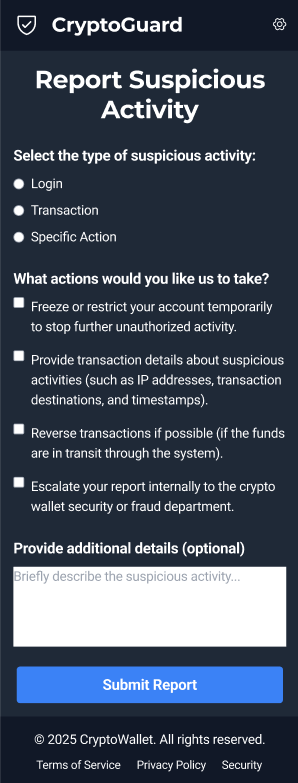}
  \caption*{(d) Report Screen}
\end{minipage}

\vspace{0.2cm}
\caption{Screens from the CryptoGuard prototype dashboard}
\label{fig.cryptoguard-screens}
\end{figure*}

The second screen, Fig. 1(b), shows the detailed login activity. In this section, users can view their recent login attempts. Here, any unusual patterns will be highlighted with red alerts. Every flagged item has a "Report" link that allows the user to take action. The third screen, Fig. 1(c), is all about transactions. It highlights flagged transfers that may indicate unauthorized or malicious activity.

The fourth screen, Fig. 1(d), shows the reporting interface. Here, users can pick the type of incident they are dealing with, whether it is a login issue or a transaction problem. They can then choose from a few follow-up options, such as freezing their account, providing more details, reversing a transaction, or escalating the case. After submission, users will be taken back to the dashboard.

Together, these design elements illustrate how the dashboard could support user decision-making in the context of cryptojacking threats. Although no user testing has been conducted yet, the prototype reflects usability-oriented choices informed by HCI literature and anticipates real-world use scenarios. This design-driven result will serve as the foundation for future evaluation.

\section{Discussion}
The design of CryptoGuard has several implications for how security dashboards can better support non-technical users in acting decisively when faced with cryptojacking threats. The core challenge is decision support: when a potential threat is flagged, what information does the user need to see, and what actions should they take in that moment to act with confidence? Our approach answers this with three deliberate choices: plain-language alerts that strip away jargon, progressive detail available only when the user requests it, and single-click actions to contain or escalate an incident. These design moves were shaped not only by HCI heuristics but also by patterns documented in prior wallet security research, where unclear alerts and overloaded screens often led users to dismiss or misinterpret warnings. These choices translate technical detection output into clear and actionable steps that a person can follow without hesitation.

\subsection{Building Trust Through Interface Design}
A central part of this design philosophy is trust. Security tools succeed or fail not only on detection accuracy, but on whether users believe the system's guidance and feel equipped to act on it. CryptoGuard's interface was deliberately structured to demonstrate that support is present at every stage, with clear labels, logically placed actions, and status feedback that reassures the user that their choices are making an impact. Trust is crucial in cryptojacking contexts, where many users may not even realize such attacks are possible. A well-structured interface closes that awareness gap by pairing clear risk communication with tangible next steps. Even though the AI engine is currently simulated, the design conveys what a real detection model would offer, removing the "black box" feeling that too often alienates non-expert users.

\subsection{From Design Concept to Implementation}
By grounding the work in human-centered design principles at this early stage, we establish an interface framework that can effectively guide and support users, regardless of the technical sophistication behind it. This positions CryptoGuard for future testing and technical integration while maintaining the core goal: enabling fast and confident responses to cryptojacking threats.

\section{Limitations and Future Work}
This initial development phase was deliberately limited to the design layer, refining the "last mile" of security, the point where detection results meet human decision-making, before introducing backend complexity. No human subjects were involved at this stage.
The current prototype is a conceptual artifact created to test UI/UX hypotheses and ensure that future technical integration builds on an interface that users already find clear, actionable, and trustworthy. This sequencing, refining the user experience before backend integration, is intentional, ensuring that technical components enhance rather than complicate the user's ability to respond quickly and accurately.

\section{Conclusion}

This paper presents CryptoGuard as a user-centered design concept for an AI-based dashboard designed to detect cryptojacking in cryptocurrency wallets. By framing it as a design concept, the scope is transparent and consistent with the early development stage, while laying the groundwork for future usability testing and AI backend integration. The proposed conceptual prototype aims to assist users in monitoring their transaction activities, identifying suspicious behavior, and enabling them to take action directly within the wallet interface. With further development and integration, CryptoGuard has the potential to significantly enhance the user experience and mitigate cryptojacking risks in cryptocurrency wallets, combining robust detection capabilities with a design that allows non-technical users to navigate with confidence.

\bibliographystyle{IEEEtran}  
\bibliography{references}

\end{document}